\documentclass[aps,prl,amsfonts,amssymb,amsmath,mathrsfs,nofootinbib,reprint,showpacs,fixltx2e,subfig,usegraphicx,twocolumn]{revtex4}
\usepackage{graphicx}
\usepackage[font=small,labelfont=bf,textfont=footnotesize,justification=raggedright,singlelinecheck=true]{caption}
\usepackage{subfig}
\usepackage{aas_macros}
\usepackage{mathrsfs}
\usepackage{comment}
\usepackage{appendix}
\usepackage{multirow}
\usepackage{url}
\newcommand{\incgraph}[3]{\includegraphics[angle=#1, width=#2\textwidth]{#3}}

\begin{document}

\title{Using Swarm Intelligence To Accelerate Pulsar Timing Analysis}

\author{Stephen R. Taylor}
\email[email: ]{staylor@ast.cam.ac.uk}
\author{Jonathan R. Gair}
\email[email: ]{jgair@ast.cam.ac.uk}
\affiliation{Institute of Astronomy, Madingley Road, Cambridge, CB3 0HA, UK }

\author{L. Lentati}
\email[email: ]{ltl21@cam.ac.uk}
\affiliation{Astrophysics Group, Cavendish Laboratory, JJ Thomson Avenue, Cambridge, CB3 0HE, UK}



\date{\today}

\begin{abstract}
We provide brief notes on a particle swarm-optimisation approach to constraining the properties of a stochastic gravitational-wave background in the first International Pulsar Timing Array data-challenge. The technique employs many computational-agents which explore parameter space, remembering their most optimal positions and also sharing this information with all other agents. It is this sharing of information which accelerates the convergence of all agents to the global best-fit location in a very short number of iterations. Error estimates can also be provided by fitting a multivariate Gaussian to the recorded fitness of all visited points.
\end{abstract}


\maketitle

\section{Introduction}
The first direct detection of gravitational-waves (GWs) will be a major step forward in astrophysics. There are plans in place which will permit detector-coverage over a huge range of GW frequencies by the late 2020s. Advanced ground-based interferometers, such as AdLIGO \citep{AdvLIGO}, AdVirgo \citep{AdvVirgo} and KAGRA \citep{kagra2012}, should be operating at design sensitivity by the end of this decade, and will be sensitive in the range $\sim1-10^3$ Hz. It is hoped that a space-based interferometer with arm-lengths of $\sim10^9$ m, such as eLISA/NGO \citep{elisa-ngo}, will be operable by the end of the 2020s, and sensitive in the range $\sim 0.1-100$ mHz.

If a GW crosses the path of an electromagnetic (EM) pulse it can induce a frequency shift of the signals which will be dependent on the amplitude of the two GW polarisations, the angle between the GW-source and the EM-source, as well as the sky-location of the EM-source \citep{sazhin-1978,burke-1975,estabrook-1975,detweiler-1979}. Thus \citet{sazhin-1978} and \citet{detweiler-1979} independently described how low-frequency ($1-100$ nHz) GWs could be detected via their influence on the arrival-times of signals from precisely timed pulsars. This is made feasible by the often sub-$\mu$s level of timing-precision achieved through the measurements of millisecond pulsar-signal time-of-arrivals (TOAs). In this case, the range of frequency-sensitivity is set by the observational time-span ($f_{\rm low}\sim1/T$) and the cadence ($f_{\rm high}\sim1/(2\Delta T)$). 

``Pulsar timing arrays'' (PTAs) \citep{foster-backer-1990} allow us the opportunity to use the Milky Way as a kpc-scale GW-detector, as tens of Galactic millisecond pulsars are observed over several years. The International Pulsar Timing Array (IPTA) \citep{ipta-site} consortium combines the efforts of the European Pulsar Timing Array (EPTA) \citep{epta-site}, the North American Nanohertz Observatory for Gravitational Waves (NANOGrav) \citep{nanograv-site}, and the Parkes Pulsar Timing Array (PPTA) \citep{ppta-site}. It recently initiated the first ``IPTA Data Challenge'' \citep{first-ipta-challenge}; the aim of this challenge was to test new and existing algorithms for the purpose of constraining the properties of a background of GWs using PTAs.

It is not only GWs which may induce deviations of the TOAs. The dominant perturbation is caused by the deterministic spindown of the pulsar itself, as its rotational energy is extracted to power the EM outflow. There are also stochastic contributions to the deviations caused by a variety of sources, including clock noise, receiver noise and variations of the dispersion measure of the intervening interstellar medium. These effects must be accounted for and removed from the TOAs to produce the timing residuals, which contain the influence due to all unmodelled phenomena, including GWs. While there is a rich literature on the subject of the detection of single GW-sources using pulsar-timing (e.g.,\ \citep{sesana-vecchio-volonteri-2009,babak-sesana-2012,sesana-vecchio-2010,lee-wex-2011,sesana-vecchio-colacino-2008,petiteau-ga-2012}), the holy grail of PTAs is a stochastic gravitational-wave background (GWB).

An isotropic, stochastic GWB may consist of a superposition of many single sources which are not individually resolvable. The largest contribution will likely be from a background induced by a cosmological population of inspiraling supermassive-black-hole-binary (SMBHB) systems, with typical masses $\sim 10^4 - 10^{10}$ M$_{\odot}$. 

The fractional energy-density of the Universe in a GW-background is usually given as,

\begin{equation}
\Omega_{\rm GW}(f) = \frac{1}{\rho_{\rm{crit.}}}\frac{d\rho_{\rm GW}(f)}{d(\ln f)} = \frac{\pi}{4}f^2h_c(f)^2,
\end{equation}

where $f$ is the observed GW-frequency, $\rho_{\rm{crit.}}$ is the energy-density required for a flat Universe, and $h_c(f)$ is the characteristic strain of the GW-background in a frequency interval centred at $f$.

The characteristic strain spectrum of a GW-background resulting from inspiraling binary systems is approximately $h_c(f)\propto f^{-2/3}$ \citep{begelman1980,phinney2001,jaffe-backer-2003,wyithe-loeb-2003}. We can approximate the characteristic strain spectrum of a GW-background from other sources as a power-law also. Some measurable primordial background contributions may have a power-law index of $-1$ \citep{grishchuk-1976,grishchuk-2005}, while the background from decaying cosmic strings \citep{vilenkin-1981a,vilenkin-1981b,olmez-2010,sanidas-2012} may have $-7/6$ \citep{damour-vilenkin-2005}. For most models of interest, we can describe an isotropic, stochastic GW-background by \citep{jenet-2006},

\begin{equation}
h_c(f) = A\left(\frac{f}{\rm{yr}^{-1}}\right)^{\alpha}.
\end{equation}

This characteristic strain spectrum is related to the one-sided power spectral density of the induced timing residuals by,

\begin{equation}
S(f) = \frac{1}{12\pi^2}\frac{1}{f^3}h_c(f)^2 = \frac{A^2}{12\pi^2}\left(\frac{f}{\rm{yr}^{-1}}\right)^{-\gamma}\rm{yr}^3,
\end{equation} 

where $\gamma\equiv3-2\alpha$.

\citet{hellings-downs-1983} developed a simple cross-correlation technique for pulsars affected by the same stochastic, isotropic GWB, showing that the cross-correlation of the induced timing-residuals has a distinctive angular signature dependant only on the angular-separation of the pulsars. This angular correlation is given by,

\begin{equation} \label{eq:hell-down}
\zeta_{ab} = \frac{3}{2}x\ln(x) - \frac{1}{4}x + \frac{1}{2} + \frac{1}{2}\delta_{ab},
\end{equation}

where $x = (1-\cos\theta_{ab})/2$, and $\theta_{ab}$ is the angular separation of the pulsar sky-locations.

We define the power-spectral density of uncorrelated red-timing noise affecting the pulsar TOAs as,

\begin{equation}
S(f) = N_{\rm red}^2\left(\frac{1}{1\text{yr}^{-1}}\right)\left(\frac{f}{1\text{yr}^{-1}}\right)^{-\gamma_{\rm red}}
\end{equation}

\section{Pulsar Timing Analysis}

Observations of pulsars lead to measurements of the pulse TOAs. The emission-time of a pulse is given in terms of the observed TOA by \citep{tempo2-1,tempo2-2},

\begin{equation}
t_{\rm em}^{\rm psr} = t_{\rm arr}^{\rm obs} - \Delta_{\odot} - \Delta_{\rm IS} - \Delta_{\rm B},
\end{equation}

where $\Delta_{\odot}$ is the transformation from the site TOAs to the Solar-system barycentre, $\Delta_{\rm IS}$ accounts for the delaying-effects as the pulse propagates through the interstellar medium, and $\Delta_{\rm B}$ converts to the pulsar-frame for binary pulsars.


In the first IPTA data challenge \citep{first-ipta-challenge} the raw data is in the form of pulsar parameter files (``.par'') and timing files (``.tim''). The parameter file contains first estimates of the pulsar timing-model parameters; these parameters describe deterministic contributions to the arrival times. The vector of measured arrival times will be composed of a deterministic and a stochastic contribution (from time-correlated stochastic signals which are modelled by a random Gaussian process),

\begin{equation}
\vec{t}^{\rm{arr}} = \vec{t}^{\rm{det}} + \delta \vec{t}^{\rm{rgp}}.
\end{equation}

The stochastic process has auto-correlation,

\begin{equation}\label{eq:pre-fit-stoch-cov}
C_{ij}=\langle\delta t_i^{\rm{rgp}}\delta t_j^{\rm{rgp}}\rangle, 
\end{equation}

where the elements of the covariance matrix are parametrised by a set of parameters, $\vec\phi$. Using the Wiener-Khinchin theorem, we can then define the auto-correlation as the Fourier transform of the power spectral density,

\begin{equation}
C(\tau) = \int_0^{\infty}S(f)\cos(f\tau)df,
\end{equation}
where $\tau=2\pi\vert t_i - t_j \vert$, and $S(f)$ is the power spectral density of the time-series $\delta \vec{t}^{\rm{rgp}}$. A closed-form expression for the auto-correlation of a time-series influenced by an underlying power-law PSD is given in \citet{van-haasteren-limits-2011}, and is used in the following.

\subsection{Processing raw arrival-times}

The ``.par'' and ``.tim'' files are fed to the \textsc{Tempo2} software package \citep{tempo2-1,tempo2-2,tempo2-3} which processes the raw arrival-times. A vector of ``pre-fit'' timing-residuals are computed using first guesses of the ``$m$'' timing-model parameters from the ``.par'' files i.e.,\ $\beta_{0,i}\rightarrow\beta_{i}$. This first guess is usually precise enough so that a linear approximation can be used in the TOA fitting procedure. In this linear approximation, the timing-residuals are described by,

\begin{equation}
\delta \vec{t} = \delta \vec{t}^{\rm{prf}} + M\vec\xi,
\end{equation}
where $\delta \vec{t}^{\rm{prf}}$ are the pre-fit timing-residuals (length $n$), $\vec\xi$ is the vector of deviations from the pre-fit parameters (length $m$) defined as $\xi_a = \beta_a - \beta_{0,a}$, and $M$ is the $(n\times m)$ ``design-matrix'', describing how the residuals depend on the timing-model parameters. \textsc{Tempo2} does not take into account the possible time-correlated stochastic signal in the TOAs, so will perform a weighted least-squares fit for the timing-model parameters. Hence it is possible that some of the time-correlated stochastic signal is absorbed in this fitting procedure, which is undesirable.

The \textsc{Tempo2} analysis provides output-residuals and the design matrix. The design matrix describes the dependence of the timing residuals on the timing-model parameters. The output-residuals form the input data vector for further study. 

\subsection{Generalised least-squares (GLS) estimator of stochastic and deterministic parameters}

We now want to re-process the \textsc{Tempo2} output-residuals to take into account the correlated stochastic signal affecting the pulse arrival times. We relate the \textsc{Tempo2} output-residuals to the stochastic contribution to the residuals by a further fitting process,

\begin{equation}
\delta \vec{t} = \delta \vec{t}^{\rm{rgp}} + M\vec\xi,
\end{equation}

where, in this case, $\delta \vec{t}$ refers to the output-residuals from \textsc{Tempo2}. 

We re-fit the timing-model for each pulsar, taking into account the possible contribution from a time-correlated stochastic process with covariance matrix $C$. This covariance matrix may contain contributions from the GWB, white-noise from TOA-errors, and possibly red-timing noise which is uncorrelated between different pulsars. The likelihood of measuring post-fit residuals, $\delta \vec{t}$, with linear parameters $\vec\xi$ and stochastic parameters, $\vec\phi$, is,

\begin{align}
\mathcal{L}(\delta\vec{t}\vert\vec{\xi},\vec{\phi}) =& \frac{1}{\sqrt{(2\pi)^n{\rm{det}} C}}\times\nonumber\\
&\exp{\left(-\frac{1}{2}\left(\delta\vec{t} - M\vec{\xi}\right)^{T}C^{-1}\left(\delta\vec{t} - M\vec{\xi}\right)\right)}.
\end{align}

This likelihood expression is effectively a GLS estimator, and is the basis for the framework used in these notes to study the first IPTA data challenge.

If we assume flat priors on the timing-model parameters then these parameters can be analytically marginalised over. The posterior distribution marginalised over timing-model parameters is \citep{van-haasteren-2011},

\begin{equation}\label{eq:long-vh-marge}
P(\vec\phi\vert\delta\vec{t})\propto\frac{1}{\sqrt{{\rm{det}}C\times{\rm{det}}(M^{T}C^{-1}M)}}\exp{\left(-\frac{1}{2}\delta\vec{t}^{T}C'\delta\vec{t}\right)},
\end{equation}

where $C' = C^{-1} - C^{-1}M\left(M^{T}C^{-1}M\right)^{-1}M^{T}C^{-1}$. When dealing with large datasets and many pulsars, $C'$ naturally involves the multiplication and inversion of high dimensional matrices.

Expression (\ref{eq:long-vh-marge}) can be written more compactly and in a way which is slightly faster to compute \citep{van-haasteren-levin-2012}:

\begin{align}\label{eq:vh-marg}
P(\vec\phi\vert\delta\vec{t})&=\pi(\vec\phi)\times\frac{1}{\sqrt{(2\pi)^{n-m}{\rm{det}}(G^{T}CG)}} \nonumber\\
&\quad\exp{\left(-\frac{1}{2}\delta\vec{t}^{T}G\left(G^{T}CG\right)^{-1}G^{T}\delta\vec{t}\right)},
\end{align}

where $G$ is the matrix of the final $(n-m)$ columns of the matrix $U$ in the SVD of the design matrix, $M=U\Sigma V^*$. The matrix $G$ can be pre-computed and stored in memory for use in each likelihood calculation. 

Equation (\ref{eq:vh-marg}) provides a robust, unbiased Bayesian framework for the search for correlated signals in PTAs, and is used with uniform priors as the optimisation function in the following section.

\section{Particle Swarm Optimisation}
\begin{table*}
\caption{\label{tab:pso-error-table}The globally optimum positions and associated $1\sigma$ errors as found by the PSO algorithm. No values are given for the red-noise parameters in \textsc{Open}3 since $N_{\rm red}$ was consistent with zero and $\gamma_{\rm red}$ was unconstrained.}
\begin{ruledtabular}
\begin{tabular}{c | c c c c c c c c c}
& \multicolumn{2}{c}{\textsc{Open}1} & \multicolumn{2}{c}{\textsc{Open}2} & \multicolumn{4}{c}{\textsc{Open}3} \\
& $A$ / $10^{-14}$ & $\gamma$ & $A$ / $10^{-14}$ & $\gamma$ & $A$ / $10^{-14}$ & $\gamma$ & $N_{\rm red}$ / ns & $\gamma_{\rm red}$\\
\hline
True & $5$ & $4.33$ & $5$ & $4.33$ & $1$ & $4.33$ & $10.1$ & $1.7$ \\
ML & $4.81$ & $4.41$ & $5.38$ & $4.33$ & $1.21$ & $4.07$ & -- & --\\
$\Delta\theta$ & $0.179$ & $0.0821$ & $0.256$ & $0.0876$ & $0.121$ & $0.155$ & -- & --\\
\end{tabular}
\end{ruledtabular}
\end{table*}

Particle swarm optimisation (PSO) is a method of finding the global maximum/minimum of non-linear functions using a swarm methodology. This technique originally developed out of a social metaphor \citep{kennedy-eberhart1995,shi-eberhart1998}, attempting to describe the dynamics of birds flocking, fish schooling and, more generally, the theory of swarms.

Swarm algorithms are becoming more popular in astronomy \citep{rogers-fiege2011,skokos2005,prasad-souradeep2012}, and, more specifically, in gravitational-wave astronomy \citep{wang-mohanty2010,gair-porter2009}. However as far as we are aware, the problem has not yet been applied to searches for a stochastic gravitational wave background with PTAs. An important caveat is that PSO will return a \textit{global} best-fit solution, but not error bars. The sampling of parameter space by particles will not necessarily be representative of the likelihood/posterior surface. 

A set of ``particles'' (computational agents or cores) are driven by ``cognition'' and ``social'' factors to explore a parameter space by carrying out random walks. Particles remember their past positions and values of the fitness/optimisation function, and share this information with all other particles. Each particle then carries out a random walk through parameter space, also experiencing a tendency to move toward their personal optimal location and the global optimal location.

First we define the terminology of PSO algorithms, using the notation of \citet{prasad-souradeep2012}. We have particles (computer cores or threads) which have positions, $\vec X$, and velocities, $\vec V$, where $\vec X^j(i)$ refers to the position of particle $j$ at iteration $i$. The fitness/optimisation function, $\mathcal{F}$, is used for searching for a global maximum/minimum, where $\mathcal{F}^j(i)$ refers to the fitness of particle $j$ at iteration $i$. We either try to maximise $\mathcal{F}=\ln P$, or minimise $\mathcal{F}=\chi^2_{\text{eff}}\equiv-2\ln P$. In the following we maximise $\ln P$.

$\texttt{Pbest}$ is the maximum value of the fitness function for particle $j$ up to the present iteration, $N$.

\begin{equation}
\texttt{Pbest}^j=\text{Max}\left\{\mathcal{F}^j(i),i=0,\ldots,N\right\},
\end{equation}

where the location of $\texttt{Pbest}^j$ is given by the vector $\vec{\texttt{P}}^j$,

\begin{equation}
\vec{\texttt{P}}^j = \vec{X}^j(i)\quad\text{if }\mathcal{F}^j(i)=\texttt{Pbest}^j.
\end{equation}

\texttt{Gbest} is the largest of the \texttt{Pbest} values among all particles. This value only changes when a particle finds a new position which is more optimal than any position any particle has ever visited.

\begin{equation}
\texttt{Gbest}=\text{Max}\left\{\texttt{Pbest}^j,j=0,\ldots,N_p\right\},
\end{equation}

where $N_p$ is the number of particles. The location of \texttt{Gbest} is given by vector $\vec{\texttt{G}}$,

\begin{equation}
\vec{\texttt{G}} = \vec{X}^j(i)\quad\text{if }\texttt{Pbest}^j=\texttt{Gbest}.
\end{equation}

\subsection{Particle dynamics}

The particle positions and velocities are updated according to,

\begin{align}
\vec{X}^j(i+1) =& \vec{X}^j(i) + \vec{V}^j(i+1),\nonumber\\
\vec{V}^j(i+1) =& w\vec{V}^j(i) + c_1u_1\left[\vec{\texttt{P}}^j-\vec{X}^j(i)\right] + c_2u_2\left[\vec{\texttt{G}}-\vec{X}^j(i)\right],
\end{align}

where $c_1$, $c_2$ are ``acceleration co-efficients'', $w$ is the ``inertia weight'', and $u_1$, $u_2$ are uniform random numbers in the range $[0,1]$.

The values of $c_1$, $c_2$ tune the contribution due to personal and social learning respectively. The ``$wV$'' term moves particles in a straight line, while the $c_1u_1[\ldots]$ and $c_2u_2[\ldots]$ terms accelerate particles toward the location of \texttt{Pbest} and \texttt{Gbest} respectively.

We use the values from the PSO Standard 2006 code \citep{PSO-site}; $w=1/(2\ln{(2)})=0.72$ and $c_1=c_2=0.5+\ln{(2)}=1.193$.

\subsection{Boundary conditions}

Particle velocities are capped to prevent them rapidly leaving the search space. The maximum velocity is typically set to be proportional to the search range. We adopt,

\begin{equation}
\vec{V}_{\text{max}} = 0.5\left(\vec{X}_{\text{max}}-\vec{X}_{\text{min}}\right),
\end{equation}

where the particle velocity is set to be $\pm\vec{V}_{\text{max}}$ if the proposed velocity is greater than $\vec{V}_{\text{max}}$ or less than $-\vec{V}_{\text{max}}$, respectively.

The ``reflecting wall'' boundary condition is used for the particle positions, such that a particle reverses it's velocity component perpendicular to the boundary if it tries to cross the boundary. Thus,

\begin{equation}
\vec{V}^j(i) = -\vec{V}^j(i),
\end{equation}

and the particle position is set to be $\vec{X}_{\text{max}}$ or $\vec{X}_{\text{min}}$ when the proposed position is greater than $\vec{X}_{\text{max}}$ or less than $\vec{X}_{\text{min}}$, respectively. Our search range for the GWB parameters are $\log(A)\in[-20,-12]$ and $\gamma\in[1,5]$.

\subsection{Initial conditions}

Random positions and velocities are assigned to each particle at the beginning of the algorithm,

\begin{align}
\vec{X}^j(i=0) &= \vec{X}_{\text{min}} + u\left[\vec{X}_{\text{max}}-\vec{X}_{\text{min}}\right],\nonumber\\
\vec{V}^j(i=0) &= u\vec{V}_{\text{max}},
\end{align}

where $u\in U\left[0,1\right]$.

\subsection{Termination criteria}

For the purposes of establishing a termination criterion for the algorithm, we treat particle trajectories like MCMC chains, even though the particle trajectories are coupled. Gelman-Rubin R statistics \citep{gelman-rubin-1992,brooks-gelman-1998} are used to determine when the swarm has sufficiently converged on the global best-fit location.

We use the potential scale reduction factor (PSRF), $\hat{R}$, of \citet{brooks-gelman-1998}. If $\hat{R}>1$ then the trajectories are not yet close to the true global optimum position. We run our PSO trajectories until $\hat{R}$ has been less than $1.02$ for all parameters for at least $50$ iterations. Analysis of all datasets finished in less than two hours with $80$ cores, although the global best-fit location was typically reached in less than $30$ minutes.

\subsection{Error estimates}

PSO is designed to explore the full parameter space (preventing all particles getting stuck in local maxima/minima) and then rapidly converge to the global optimal location. A binning of all points visited by particle trajectories will not reconstruct the posterior surface.

We can approximate the posterior surface close to the global best-fit point as a Gaussian,

\begin{equation}
P = P_0\exp{\left(-\frac{1}{2}\vec\Delta^TC'\vec\Delta\right)},
\end{equation}

where $\Delta_a = \left(\theta_a - \texttt{G}_a\right)/\texttt{G}_a$ for parameter $\theta_a$, and $C'=C^{-1}$.

The standard error, $\Delta\theta_a$, in parameter $\theta_a$ is given by,

\begin{equation}
\Delta\theta_a = \sqrt{C_{aa}}\times\texttt{G}_a = \sqrt{\left(C'^{-1}\right)_{aa}}\times\texttt{G}_a.
\end{equation}

We can fit a paraboloid to $\Delta \chi_{\text{eff}}^2 = -2\left(\ln P-\ln P_0\right)$ (where we take $P_0$ to be the posterior at $\vec{\texttt{G}}$), and identify the fitting co-efficients with the elements of $C'$. Hence,

\begin{equation}
\Delta \chi_{\text{eff}}^2 = \vec\Delta^TC'\vec\Delta.
\end{equation}

We limit the fitting to a subset of points, where only points within a hypersphere of radius $|\Delta_a|<0.3$, and $\Delta \chi_{\text{eff}}^2<10$ are considered. This limited the fitting to several thousand points.

\section{Conclusions}
\begin{figure*}
  \subfloat[]{\label{fig:gbest_amp_cores}\incgraph{270}{0.45}{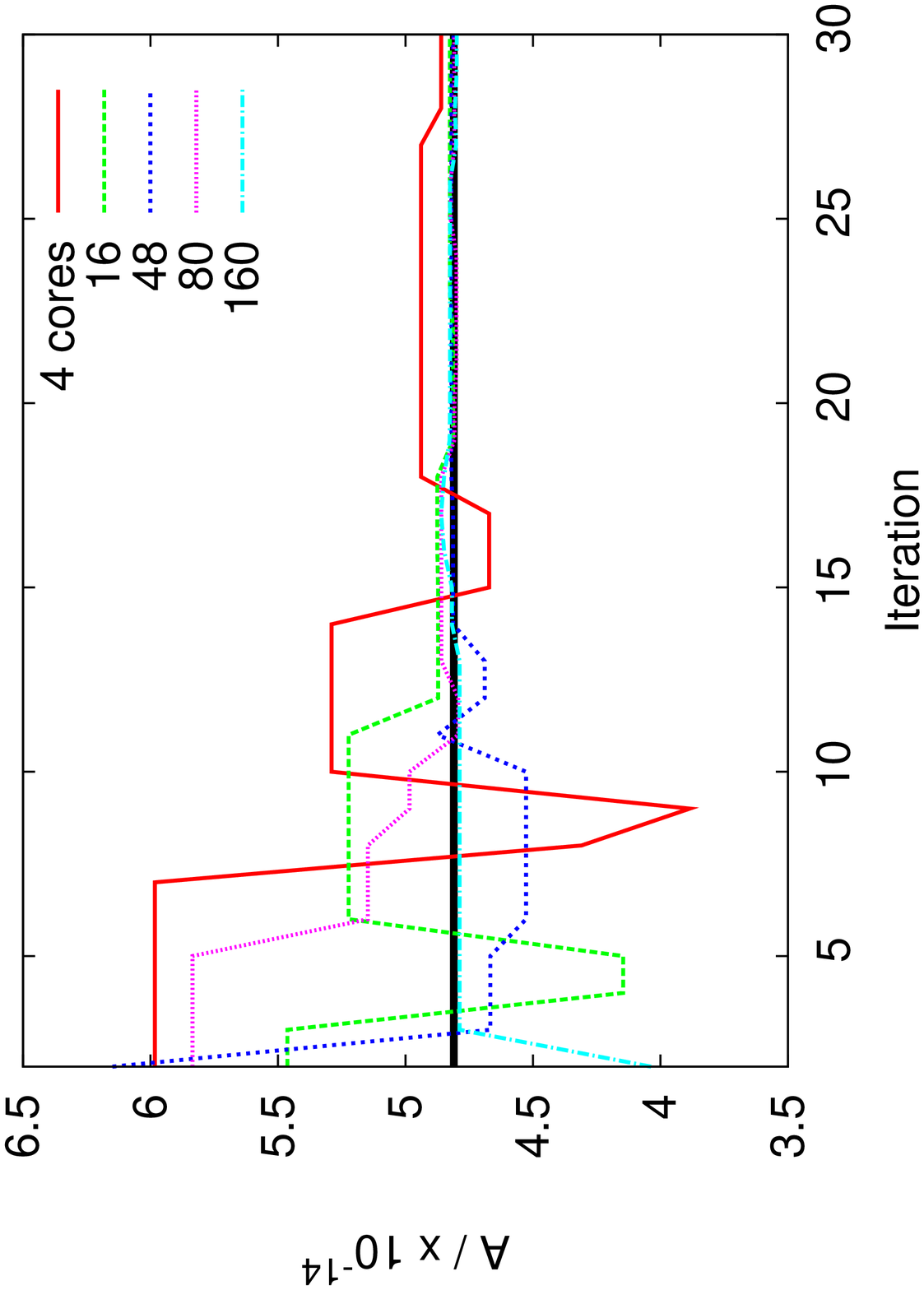}} 
  \subfloat[]{\label{fig:gbest_gamma_cores}\incgraph{270}{0.45}{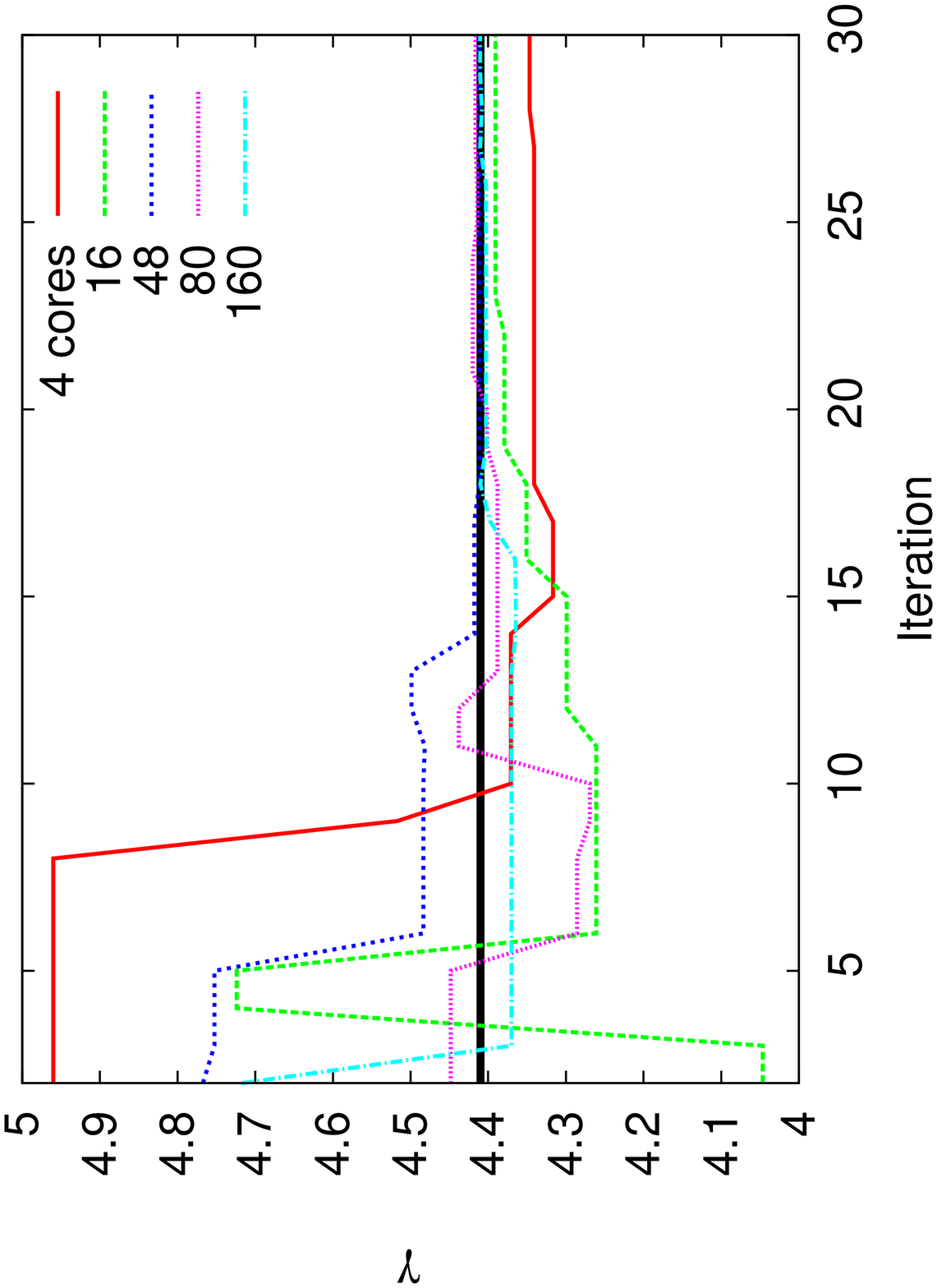}} 
  \caption{\label{fig:gbest_cores}The updating values of the global best-fit location in the parameter space of \textsc{Open}1 are compared for varying numbers of cores/threads. We show approximately the first half-hour of the analysis. The mode of the posterior surface at $(A=4.81\times10^{-14},\gamma=4.41)$ is shown as a black line in each panel. Beyond a few tens of cores, the speed-up in reaching the global best-fit location is minimal.}
\end{figure*}
The globally optimum positions found by the PSO trajectories and the associated $1\sigma$ errors for all \textsc{Open} datasets are listed in Table \ref{tab:pso-error-table}. In all cases, we have $36$ pulsars with $130$ observations each, spanning over approximately $5$ years. The white-noise in each pulsar is read-in as the TOA-error. In Figure \ref{fig:gbest_cores} we show how the global best-fit location of \textsc{Open}1 is updated with time for varying numbers of cores/threads. As confirmed with MCMC and Nested Sampling, we are dealing with a unimodal posterior surface, so beyond a few tens of cores the gains in the speed at which the global best-fit location is reached are minimal. However, if we had a multimodal surface then it would be desirable to have many PSO trajectories to aggressively search the parameter space for the \textit{global} best-fit location. We recommend a few tens of cores, not only as a compromise between speed at which the best-fit is reached and computational expenditure, but also to permit an accurate reconstruction of the posterior in the vicinity of the mode for error estimates.

In the interest of fairness, we do not give results for the \textsc{Closed} datasets here, but will update this manuscript after the challenge deadline has passed and the results are widely available. The values in both \textsc{Open} and \textsc{Closed} datasets compare very favourably with traditional stochastic sampling techniques such as MCMC and Nested Sampling. This will be discussed in an article which is in preparation by the authors and will be submitted after the deadline ends. However, the PSO error estimates may not accurately reflect confidence intervals if we have a significantly non-Gaussian posterior surface. 

\vspace{2cm}

\begin{acknowledgments}
S.R.T is supported by the STFC. J.R.G is supported by the Royal Society. L.L is supported by the STFC. We thank Rutger van Haasteren for incredibly helpful advice and discussions. This work was performed using the Darwin Supercomputer of the University of Cambridge High Performance Computing Service (http://www.hpc.cam.ac.uk/), provided by Dell Inc. using Strategic Research Infrastructure Funding from the Higher Education Funding Council for England.
\end{acknowledgments}

\bibliography{pta_refs}

\end{document}